\documentclass[10pt,preprintnumbers,amsmath,amssymb,floatfix,prl,superscriptaddress,twocolumn]{revtex4}

\usepackage{graphicx}
\usepackage{bm}
\usepackage{amsmath}
\usepackage{amssymb}
\usepackage{bbold}

\begin{document} 




\title{Haldane phase in the Hubbard model for the organic molecular compound Mo$_3$S$_7$(dmit)$_3$}

\author{C. Janani}
\email{jananichander84@gmail.com}
\affiliation{Centre for Organic Photonics and Electronics, School of Mathematics and Physics, The University of Queensland, Brisbane, Queensland 4072, Australia} 
\affiliation{Centre for Engineered Quantum Systems, School of Mathematics and Physics, The University of Queensland, Brisbane, Queensland 4072, Australia}
\author{J. Merino}
\affiliation{Departamento de F\'{i}sica Te\'{o}rica de la Materia Condensada, Condensed Matter Physics Center (IFIMAC) and Instituto Nicol\'as Cabrera, Universidad Aut\'{o}noma de Madrid, Madrid 28049, Spain}
\author{I. P. McCulloch}
\affiliation{Centre for Engineered Quantum Systems, School of Mathematics and Physics, The University of Queensland, Brisbane, Queensland 4072, Australia}
\author{B. J. Powell}
\affiliation{Centre for Organic Photonics and Electronics, School of Mathematics and Physics, The University of Queensland, Brisbane, Queensland 4072, Australia}

\begin{abstract}
We report the discovery of a correlated insulator with a bulk gap at two-thirds filling in a geometrically frustrated Hubbard model that describes the low-energy physics of Mo$_3$S$_7$(dmit)$_3$. This is very different from the Mott insulator expected at half-filling. We show that the insulating phase, which persists even for very weak electron-electron interactions ($U$), is adiabatically connected to the Haldane phase and is consistent with experiments on Mo$_3$S$_7$(dmit)$_3$. 
\end{abstract}

\maketitle

Many materials  display insulating behaviors which cannot be understood from the conventional band theory of solids.  
In contrast to band insulators, correlated insulators often have partially filled bands. Prominent examples are Mott insulators:  half-filled systems in which the onsite Coulomb repulsion \cite{Mott} between electrons, $U$, opens a gap 
and interesting magnetic properties arise. Mott physics is key to understanding strongly correlated systems such 
as the high-T$_c$ cuprate superconductors \cite{Lee,Anderson} and organic superconductors \cite{Powell}. Other examples of correlated insulators are covalent \cite{Zaanen} 
and charge transfer insulators \cite{Sarma,Merino-interplay}. Identifying new correlated insulating materials and characterizing their electronic properties 
is a fundamental challenge in condensed  matter physics and promises future applications.

\begin{figure}
 \begin{center}
 \includegraphics[width = 0.9\columnwidth]{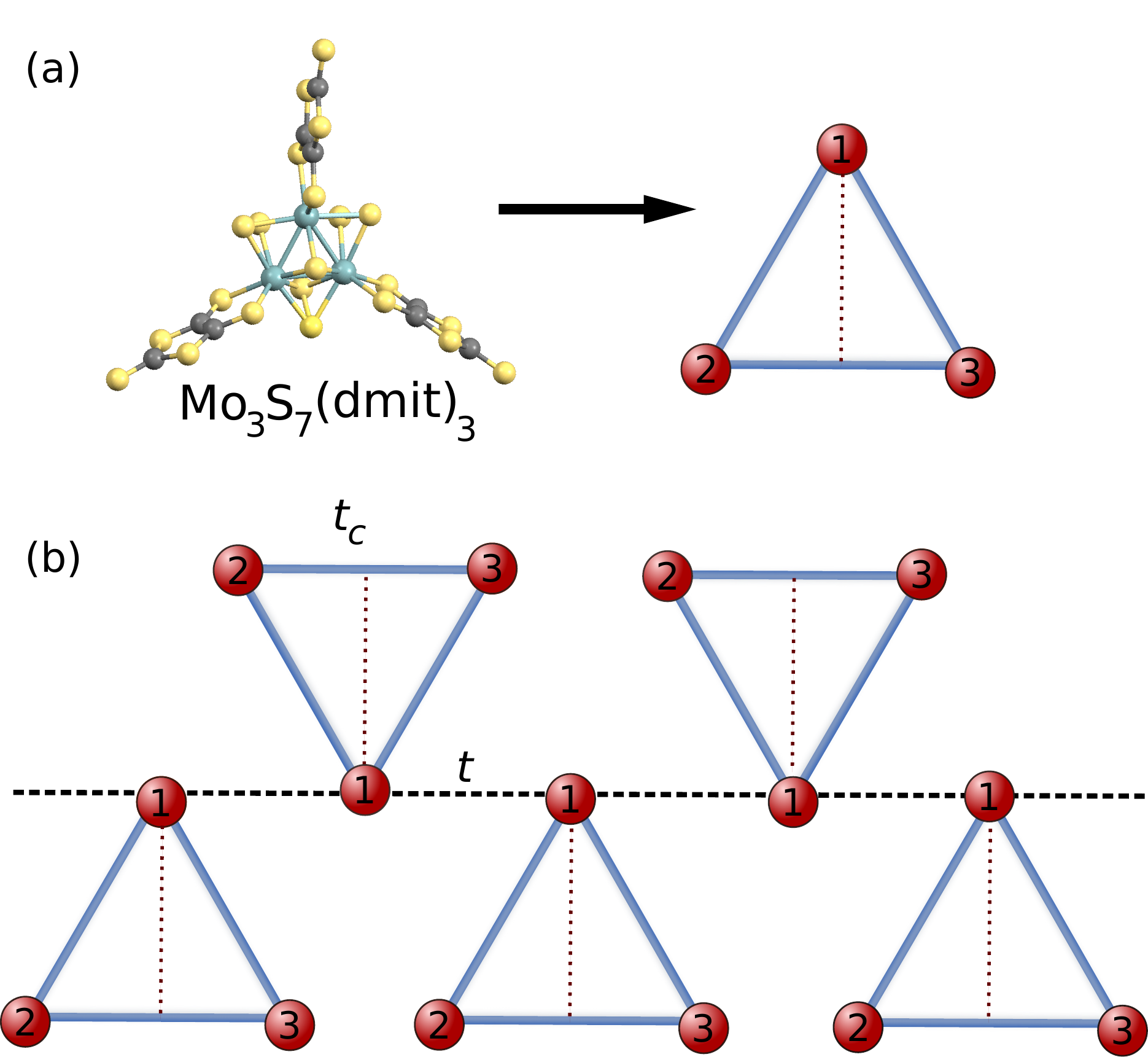}\vspace*{1cm}\\
 \includegraphics[width=0.9\columnwidth]{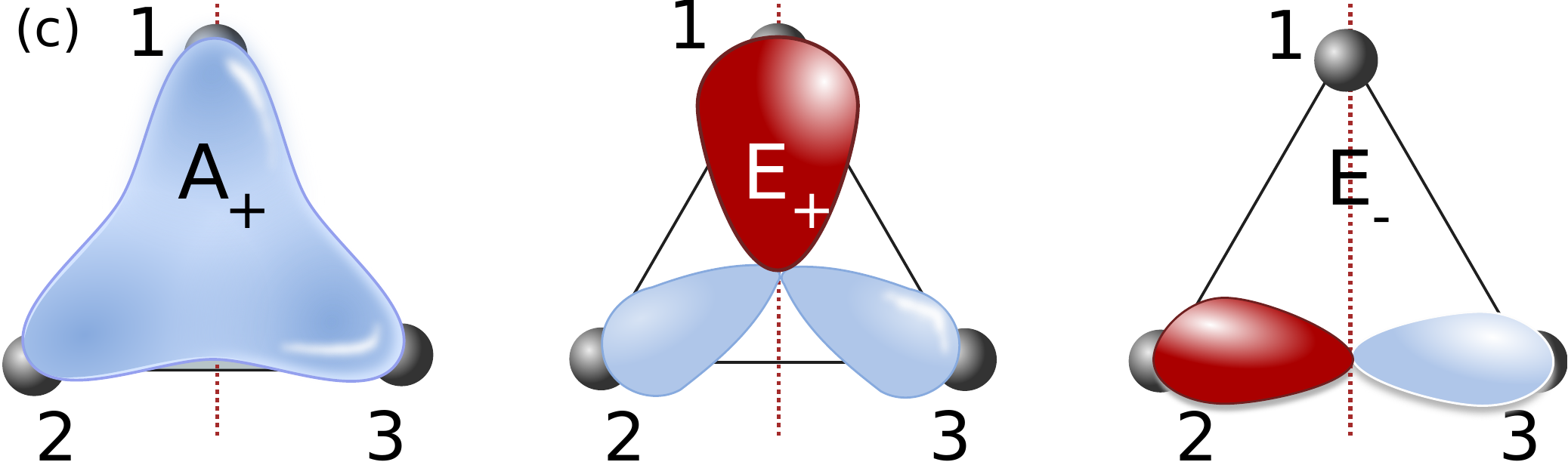}
\end{center}
\caption{(color online)
(a) The  Mo$_3$S$_7$(dmit)$_3$ molecule and its schematic representation in the Hubbard model. (b)  The triangular necklace model of  Mo$_3$S$_7$(dmit)$_3$.  (c) Sketches of the molecular orbitals, $\hat c_{iA_+\sigma}=(\hat c_{i1\sigma}+\hat c_{i2\sigma}+ \hat c_{i3\sigma})/\sqrt{3}$,  $\hat c_{iE_-\sigma}=(\hat c_{i2\sigma}-\hat c_{i3\sigma})/\sqrt{2}$, and $\hat c_{iE_+\sigma}=(2\hat c_{i1\sigma}-\hat c_{i2\sigma}-\hat c_{i3\sigma})/\sqrt{6}$,   which are  the eigenbasis when $t =U = 0$.  Different colours imply different signs. The labels $A$ and $E$ refer to the $C_{3}$ symmetry of the individual molecules and the `local parity' label ($\pm$) describes the change in phase of the orbital  on relabelling sites 2 and 3 on any single molecule, which is equivalent to reflection through the red dotted lines in panels (b) and (c).}
\label{figure:sketch}
\label{orbitals}
\end{figure}

Relatively little is known, experimentally, about Mo$_3$S$_7$(dmit)$_3$. It has a charge gap, but neither a spin gap nor long range magnetic order is observed down to 2.1~K  \cite{jacs}.
 Density functional  calculations predict that Mo$_3$S$_7$(dmit)$_3$ is a quasi-one--dimensional metal in the absence of magnetic order and a charge gap is only found when long range magnetic order is (counterfactually) assumed  \cite{jacs,dft}. On the basis of these calculations and the crystal structure of Mo$_3$S$_7$(dmit)$_3$, Llusar \emph{et al.} \cite{jacs} argued that the low energy physics is described by a classical spin model on the  `triangular necklace lattice' 
 (Fig. \ref{figure:sketch}), and showed that this model reproduces the observed temperature dependence of the magnetic susceptibility. However, neither this model nor density functional theory are able to explain why the insulating state arises in the absence of long-range magnetic order, as is found experimentally.

In this Letter we 
analyze the simplest model of interacting itinerant fermions for Mo$_3$S$_7$(dmit)$_3$, {\it viz}. the Hubbard model, on the triangular necklace lattice (Fig. \ref{figure:sketch}) at the (two-thirds) filling  [{\it i.e.}, $n=4$ electrons per triangular molecule on average].
We find a significant charge gap, but a spin gap too small to have been observed in the experiments on Mo$_3$S$_7$(dmit)$_3$ to date. Although there is no explicit Hund's rule coupling in the model Hamiltonian (Eq. (\ref{ham})) we find that, in the strong coupling limit, large molecular moments arise from a complex interplay between kinetic and interaction effects. We show that the insulating state  is adiabatically connected to the ground state of the spin-one Heisenberg model: the Haldane phase  \cite{Haldane,AKLT}.  


The Haldane phase is a key example of a symmetry protected topological (SPT) phase \cite{GuWen,Wen12}. In spin-1 chains the Haldane phase is protected by any of three symmetries: inversion, time reversal and dihedral symmetry, $D_2\cong Z_2\times Z_2$, which is equivalent to spin rotation by $\pi$ about any pair of perpendicular axes \cite{Pollmann}. That is, provided at least one of these symmetries is not explicitly broken a phase transition separates the Haldane phase from the trivial state.

Previously, Anfuso and   Rosch \cite{Anfuso} have studied a family of fermionic Hamiltonians  that extrapolate smoothly between the band insulator, the Haldane chain and the antiferromagnetic spin-$1/2$ ladder. This suggested that the Haldane phase may not be topologically distinct in fermionic systems.  Pollmann {\it et al.} \cite{Pollmann} pointed out that these models explicitly break inversion symmetry and argued that inversion symmetry could protect the topological order even in fermionic systems, but did not provide an explicit example. Interestingly, we find that in the model considered here the topologically non-trivial Haldane phase survives even in the presence of significant charge fluctuations, which suppress the magnetic moment to be significantly less than one.

   The Hamiltonian for the Hubbard model on the triangular necklace lattice is
\begin{eqnarray}
\hat H=U\sum_{i\alpha} \hat  c^{\dag}_{i\alpha\uparrow}\hat c_{i\alpha\uparrow} \hat c^{\dag}_{i\alpha\downarrow}\hat c_{i\alpha\downarrow}
-t_c\sum_{i, \alpha \neq \beta, \sigma} \hat c^{\dag}_{i\alpha\sigma} \hat c_{i\beta\sigma}\notag\\
 -t\sum_{i\sigma} \left(\hat c^{\dag}_{i1\sigma} \hat c_{(i+1)1\sigma} 
 +H. c.\right), \label{ham}
\end{eqnarray}
 where 
$\hat c^{(\dag)}_{i\alpha\sigma}$ annihilates (creates) an electron with spin $\sigma$ on the $\alpha^{th}$ site of the  $i^{th}$ molecule. 
For the  $t_c>0$ and $n=4$, the case relevant to Mo$_3$S$_7$(dmit)$_3$, the system is a topologically 
trivial metal when $U=0$.

The triangular necklace model is reminiscent of the three leg tube. The half-filled Hubbard model on this lattice has been studied at half-filling in the strong-coupling (large $U$) limit \cite{Sakai}. This model was found to display a gapped phase that can be suppressed by varying the `rung' hopping strengths  around the triangles can drive the system between different phases. However, we are not aware of any studies of this model  that considered different hopping integrals on different legs, which is the limit required to reach the triangular necklace model, or that considered 2/3-filling -- appropriate to Mo$_3$S$_7$(dmit)$_3$.

 We apply the density matrix renormalization group (DMRG) using the matrix product state (MPS) ansatz with $SU(2)$ symmetry
  \cite{Schollwock10}, keeping up to 2000 states in each  DMRG sweep, which is equivalent to $\sim9000$ states if only $U(1)$ symmetry is utilized. Except where otherwise stated, the results presented below are for a lattice size  $L= 40$  (where $L$ is the number of molecules, \emph{i.e.}, there are $3L$  sites), with  $t/t_c=0.25$. Where more appropriate we have applied infinite DMRG. Other values of $t/t_c$ give qualitatively similar results and will not be discussed at length for clarity. Whenever required we have implemented finite size  and/or finite basis set scaling.
  
\begin{figure*}
 \begin{center}
 \includegraphics[height =0.56\columnwidth,]{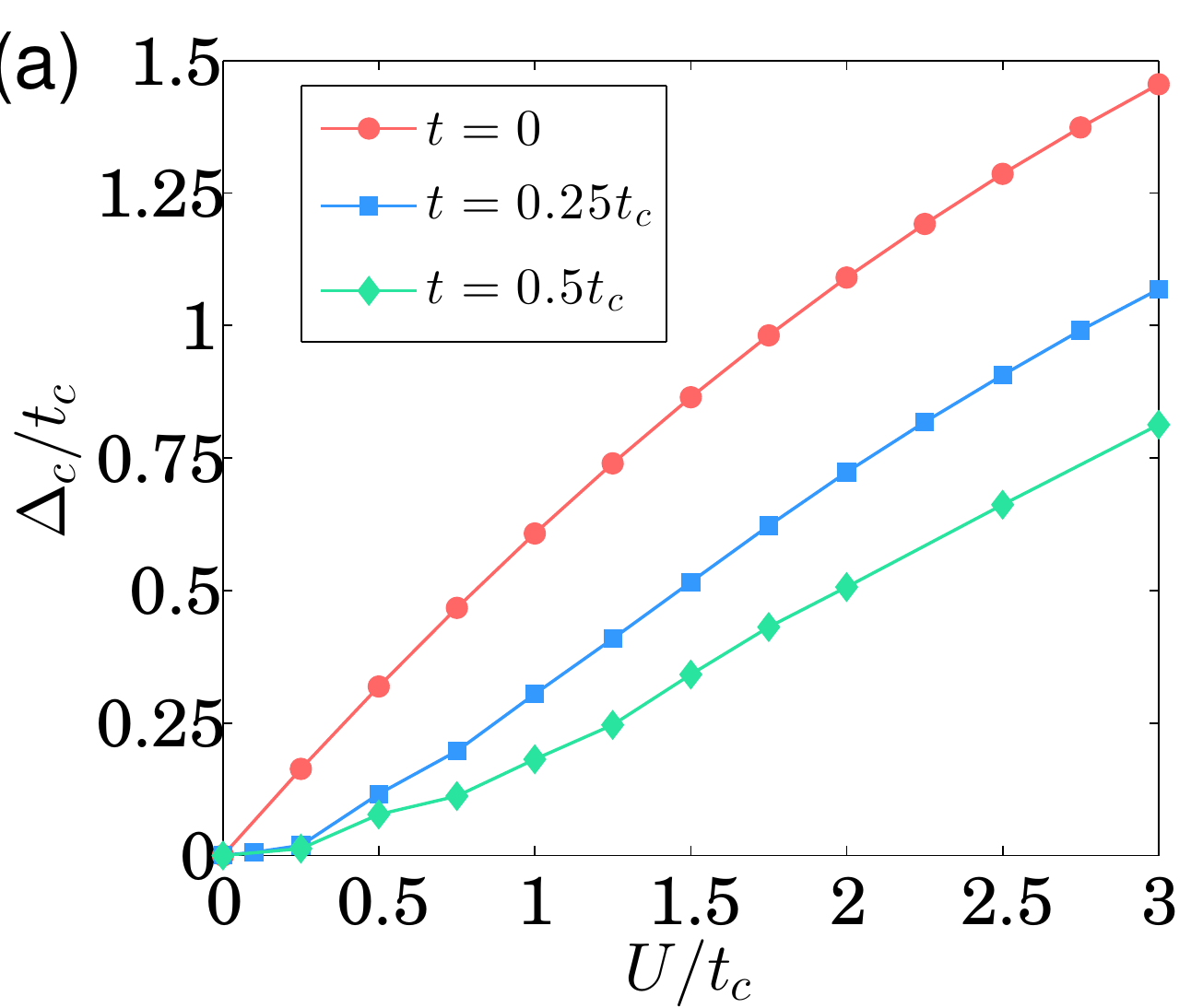}
  \includegraphics[height =0.56\columnwidth]{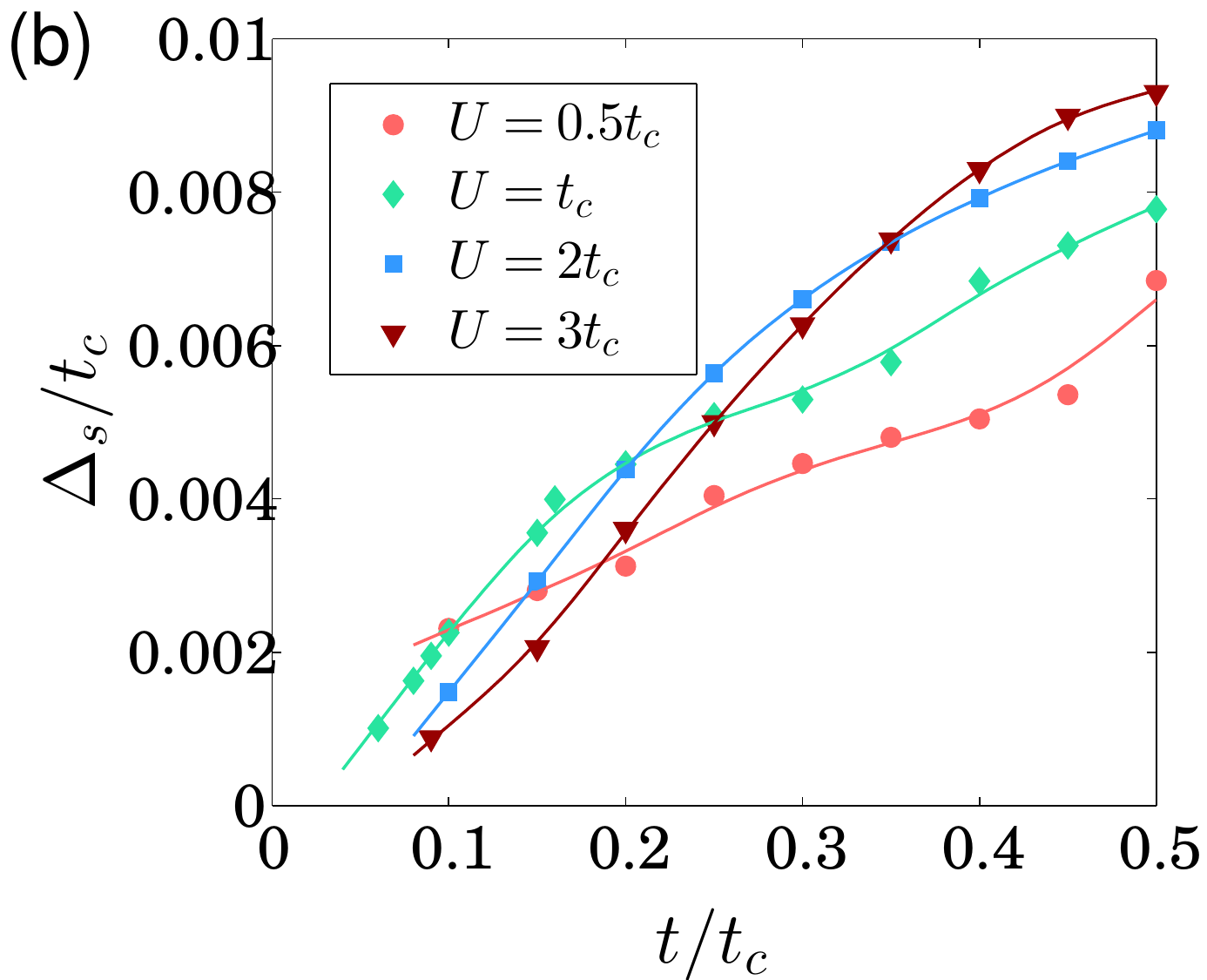}
   \includegraphics[height = 0.56\columnwidth]{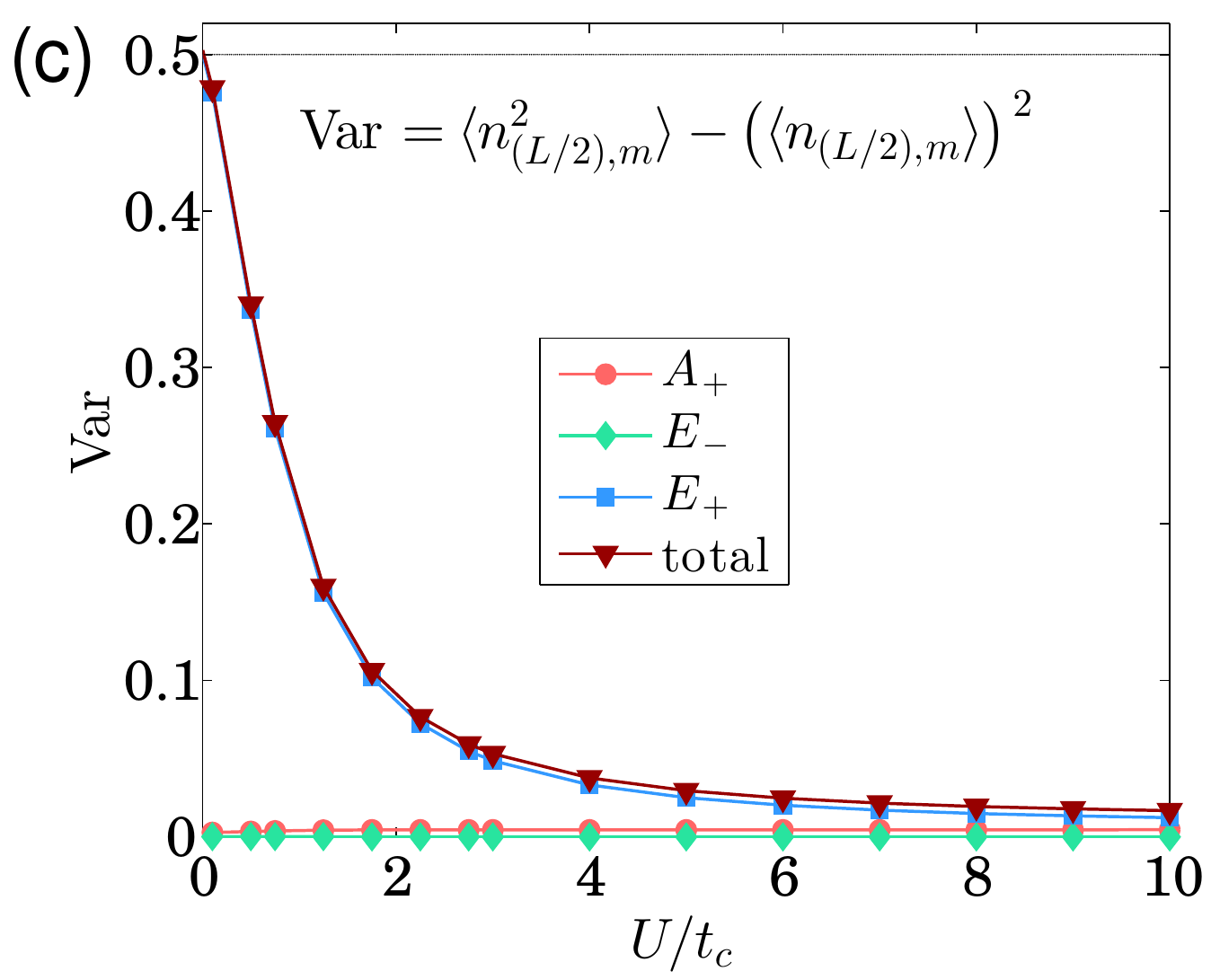}
\end{center}
\caption{(color online) 
(a) The   charge gap, $\Delta_c=[E_0(4L+2)+E_0(4L-2)-2E_0(4L)]/2$, where $E_S(N_e)$ is the energy of the  spin $S$ ground state  for $N_e$ electrons on $L$ molecules  (finite size scaled). 
\label{fig:charge-gap}
(b)   The spin gap,   $\Delta_s=E_2(4L)-E_0(4L)$  for  $L=40$ molecules,  is orders of magnitude smaller than the charge gap, $\Delta_c$. 
\label{figure:spin_gap}
(c)   The  variance in particle number in each of the molecular orbitals and  the total variance in particle number for $t=0.25t_c$. Even for small $U$ the local parity symmetry means that there are no charge fluctuations in the $E_-$ orbitals for $\langle \hat n_{iE_-}\rangle=1$. In the insulating phase $\langle\hat n_{iA_+}\rangle\lesssim2$ and $\langle \hat n_{iE_+}\rangle\gtrsim\langle \hat n_{iE_-}\rangle=1$. As the  $A_+$ orbitals are nearly-filled,  charge fluctuations in the $A_+$ orbital are significantly smaller  than  the charge fluctuations in the $E_+$ orbital. 
In all panels, curves are guides to the eye.
\label{fig:charge-fluc}
}
\end{figure*}

We find an insulating ground state for $U>0$, as is evident from the large charge gap, $\Delta_c$, shown in Fig. \ref{fig:charge-gap}a. This is surprising at two-thirds filling ($n=4$) 
and is clearly not the usual Mott insulator  expected at half-filling ($n=3$). 
As we have an average of four electrons per triangular molecule in the strong coupling limit ($U\rightarrow\infty$) one's na\"ive expectation  is   for a strongly correlated metal, with one electron per site and the remaining one-third of an electron per site free to move along the chain. Contrary to this expectation,  $\Delta_c$ continues to grow as $U$ is increased, demonstrating that the large $U$ insulating state is highly non-trivial.  For very small $U$, the charge gap becomes small and the finite size scaling is non-trivial. Nevertheless, the charge gap certainly opens at small $U$ and our numerical results do not rule out a charge gap  for any non-zero $U$.

\begin{figure*}
 \begin{center}
   \includegraphics[height =0.51\columnwidth]{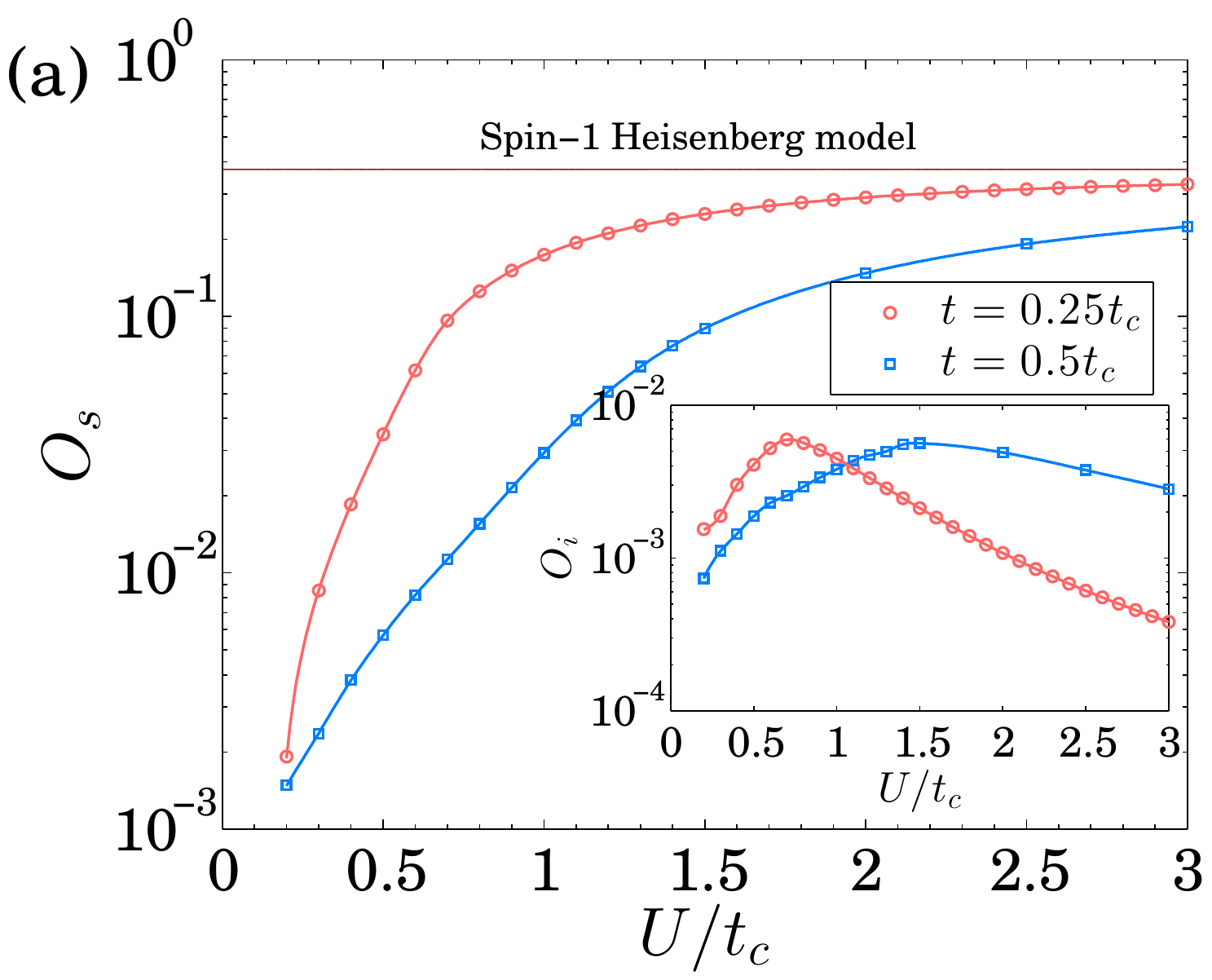}
  \includegraphics[height = 0.51\columnwidth]{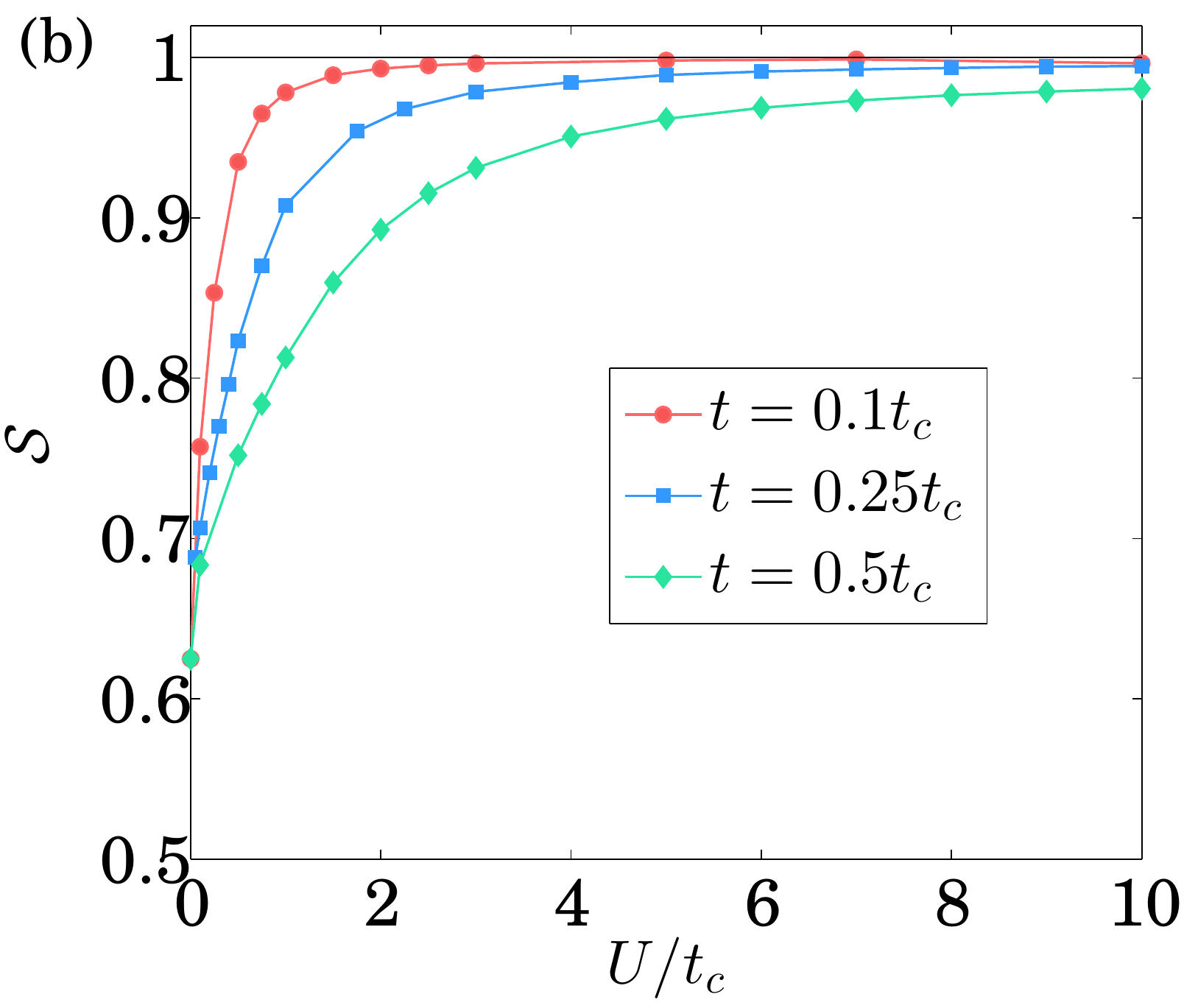}
    \includegraphics[height=0.51\columnwidth]{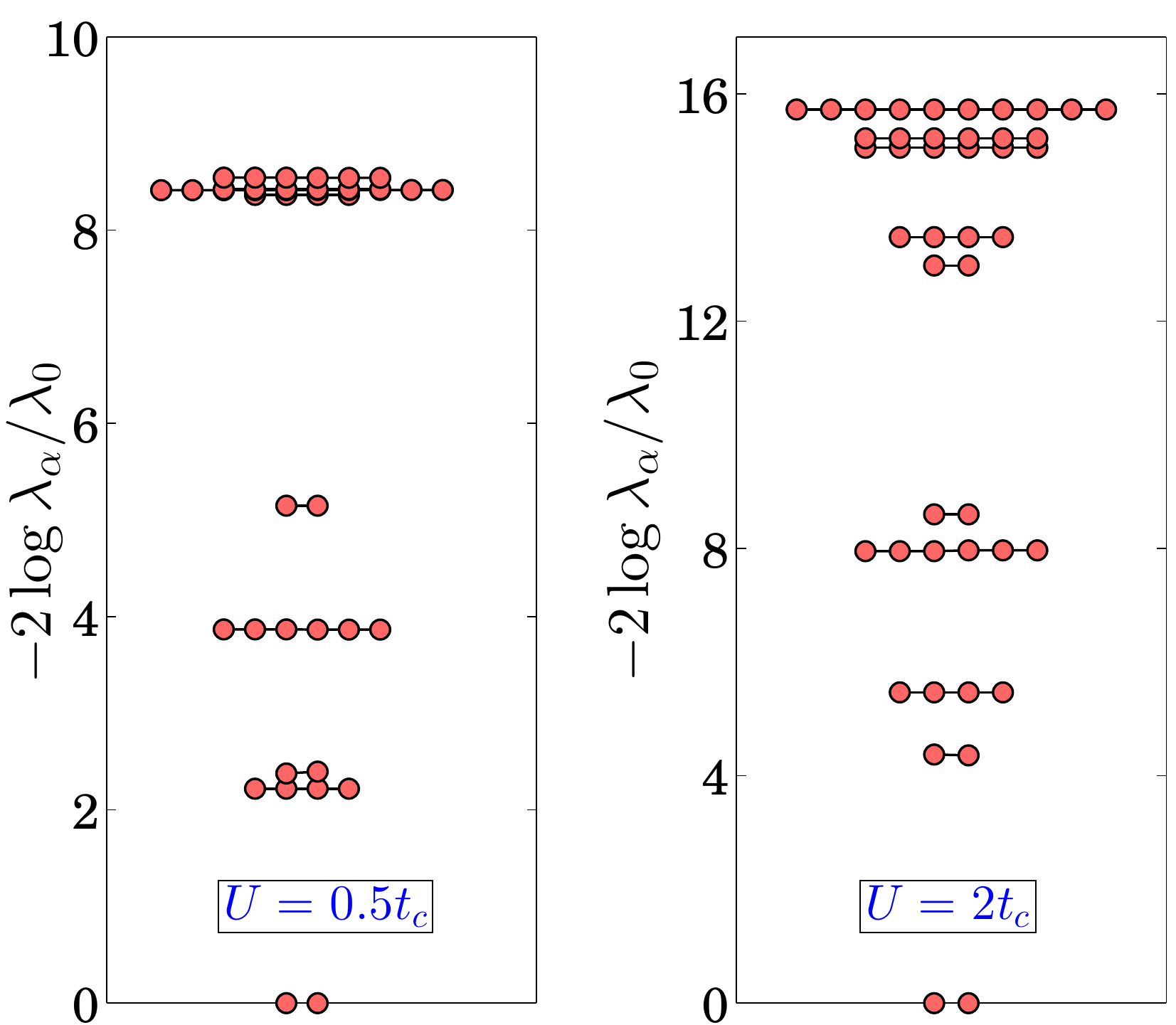}
\end{center}
\caption{(color online)
(a)   The string order parameter $O_s=\lim_{|i-j|\rightarrow\infty}\langle S^z_i\exp\left(i\pi\sum_{l=i+1}^{j-1}S^z_l\right)S^z_j\rangle$ (main panel) and $O_i=\lim_{|i-j|\rightarrow\infty}\langle \mathbb{1}_i\exp\left(i\pi\sum_{l=i+1}^{j-1}S^z_l\right)\mathbb{1}_j\rangle$  (inset; both finite basis set scaled from infinite DMRG), where $\hat{\bm{S}}_i=\sum_\alpha \hat{\bm{S}}_{i\alpha}$ is the  spin of the $i^\text{th}$ molecule, $\hat{\bm{S}}_{i\alpha}=\sum_{\sigma\sigma'}\hat c^{\dag}_{i\alpha\sigma}{\bm\tau}_{\sigma\sigma'}\hat c_{i\alpha\sigma'}$,   ${\bm\tau}_{\sigma\sigma'}$ is the vector of Pauli matrices, and $\mathbb{1}_i$ is the identity operator on the $i^\textrm{th}$ site.  
For comparison the values of $O_s$ for  spin-one Heisenberg chain  \cite{White} is shown. $O_i=0$ for the  spin-one Heisenberg chain   and the AKLT model  \cite{PollmanIdent}
\label{figure:string_order}
(b) The effective total spin per triangular molecule, $\mathcal{S}$, given by the solution of $\mathcal{S}(\mathcal{S}+1)=\langle \hat{\bm{S}}_{L/2}\cdot\hat{\bm{S}}_{L/2}\rangle$, for $L=40$.  $\mathcal{S}\rightarrow1$ as $U\rightarrow\infty$.  
\label{fig:molecular-spin} 
(c) The entanglement spectrum, {\it i.e.}, the eigenvalues of the reduced density matrix on tracing out half of the system, for $t=0.25t_c$. Degenerate data points are offset on the abscissa  for clarity. Even for small values of $U$ the entanglement spectrum has even-fold degeneracies; this is a robust signature that the SPT phase survives.
\label{fig:entanglement}
}
\end{figure*}

We also find a spin gap  (Fig. \ref{figure:spin_gap}b), which is orders of magnitude smaller than the charge gap. 
For periodic boundary conditions the ground state is unique. However, for open boundary conditions a triplet state is degenerate with the singlet ground state; these two states are separated from the remaining excitations by the spin gap. This is precisely the topologically dependent spectra that results from the ($D_2\cong Z_2\times Z_2$) symmetry of the Haldane phase  \cite{Kennedy} due to spin-1/2 edge states.
Although there is no long range magnetic order, we find a finite expectation value for the string order correlation function (Fig. \ref{figure:string_order}a) in the thermodynamic limit.
We stress that none of these phenomena are found in the  Mott insulating phase of the half-filled linear Hubbard chain, where the spin degrees of freedom form a Luttinger liquid.

In the remainder of this paper we give a simple explanation of this physics and show that the insulating phase is in the same SPT phase as the Haldane phase.
Understanding the  insulating phase is ultimately simpler if one works in the  `molecular orbital' basis, shown in Fig. \ref{orbitals}c. 
 However,  the interaction terms take a significantly more complicated form in the molecular orbital basis \cite{pert}.

It is helpful to begin by examining the strong coupling   ($U/t_c\rightarrow\infty$) limit for isolated molecules  ($t=0$). A particle-hole transformation leaves us with $n=2$ and $t_c<0$. It immediately follows from Nagaoka's theorem  \cite{Nagaoka} that the ground state is a fully polarized ferromagnet, {\it i.e.}, a triplet.  For the discussion below, it is helpful to also consider theses triplets  in the molecular orbital basis, even without making a particle-hole transformation. Firstly, we note that 
Although the Hubbard $U$ is the same on all sites, the repulsion between two electrons in an $A_+$ orbital ($U/3$) is less than the repulsion between two electrons in an $E_+$ or $E_-$ orbital ($U/2$). For four electrons in three orbitals, there must be (at least) one doubly occupied orbital; clearly in the strong coupling limit this will be the $A_+$ orbital. 
In the molecular orbital basis there is a direct exchange interaction,  $J_{E_+E_-}=-U/6$, between electrons in the $E_-$ and $E_+$ states \cite{pert}, which stabilises  the triplet, as required by Nagaoka's theorem \cite{Nagaoka}. 
Indeed, on the isolated three site cluster this argument holds for all $U>0$ and the exact solution has a triplet ground state  \cite{Merino}. Indeed it has been shown that in non-bipartite one-dimensional systems the fully polarized Nagaoka-type state is stable in a large region of parameter space away from the infinite $U$ limit \cite{Arita,Daul,Nakano}.

A non-zero intermolecular coupling ($t\ne0$)  means that the 1-sites are no longer equivalent to the 2- or 3-sites. However, the Hamiltonian still retains a `local parity' symmetry under the relabelling  of sites 2 and 3 on \emph{any} individual molecule (cf. Fig. \ref{figure:sketch}). Thus the local parity of every molecule is a constant of the motion for the full many-body wavefunction. As  $E_-$  is the {\it only} odd parity orbital, this implies that the occupation number of this orbital,  $ \hat n_{iE_-}= \sum_\sigma\hat c^\dagger_{iE_-\sigma} \hat c_{iE_-\sigma}$, is conserved modulo two. However, we found above that in the strong coupling molecular limit the ground state has exactly one electron in the $E_-$ orbital on every molecule. It follows that perturbations that do not break the local parity symmetry, such as a finite $U$ or a non-zero $t$, will not change the number of electrons in any of the $E_-$ orbitals unless they drive a phase transition. 
We find that  $\langle \hat n_{iE_-}\rangle=1$ and $\langle n_{iE_-}^2\rangle-\langle n_{iE_-}\rangle^2=0$ throughout the insulating phase (Fig. \ref{fig:charge-fluc}c),  confirming that there are no charge fluctuations in the $E_-$ orbitals.

As the $E_+$ and $A_+$ orbitals have even local parity there is no preclusion of charge fluctuations in these orbitals for finite $U$. Nevertheless, the charge gap indicates that charged excitations are confined in the insulating phase  \cite{Mott}. Thus, we see that a complex interplay of kinetic and potential effects drives the insulating phase of the two-thirds filled triangular necklace model.

We have shown previously \cite{pert} that in the molecular limit, $t/t_c\rightarrow0$, the spins on neighboring molecules are coupled by an antiferromagnetic superexchange interaction, given by $J_s=\sum_{i=0}^4{4t^2}/{[9 a_i(3t_c+\varepsilon_i)]}$ to second order \cite{foot-coef}. 
As expected from the analysis above the effective spin per molecule, ${\cal S}\rightarrow1$ in the strong coupling limit ($U/ t_c\rightarrow\infty$), see Fig. \ref{fig:molecular-spin}b. 
Thus, the low-energy physics of the two-thirds filled Hubbard model on the triangular necklace lattice in the strong coupling molecular limit is captured by the spin-one Heisenberg chain. A corollary to this is that in the strong coupling molecular limit the model is in the Haldane phase, consistent with our numerical results. 

However, as we move away from the strong coupling molecular limit an additional complication arises. The charge fluctuations in the $A_+$ and $E_+$ orbitals lead to a suppression of the effective moment on each molecule, {\it cf.} Fig \ref{fig:molecular-spin}b. As the physics of the Heisenberg chain  is  strongly dependent on the magnitude of the spin it is important to ask, particularly for small $U$, whether the charge fluctuations are sufficient to move the system out of the Haldane phase  \cite{Pollmann,Anfuso}. 

In Fig. \ref{figure:string_order}a we plot the usual string order parameter for the Haldane phase, $O_s=\lim_{|i-j|\rightarrow\infty}\langle S^z_i\exp\left(i\pi\sum_{l=i+1}^{j-1}S^z_l\right)S^z_j\rangle$, and $O_i=\lim_{|i-j|\rightarrow\infty}\langle \mathbb{1}_i\exp\left(i\pi\sum_{l=i+1}^{j-1}S^z_l\right)\mathbb{1}_j\rangle$. In spin-one models $O_s\ne0$ and $O_i=0$ in the Haldane phase, whereas $O_s=0$ and $O_i\ne0$ in the trivial phase \cite{PollmanIdent}. In the Hubbard model we find that both $O_s\ne0$ and $O_i\ne0$. Indeed, for small $U, t$ we find that $O_i>O_s$. Furthermore,  in spin-one models one can define  \cite{PollmanIdent} a projective representation of $Z_2\times Z_2$ by
\begin{eqnarray}
\sum_{\sigma'}R^\alpha_{\sigma\sigma'}A^{\sigma'}=e^{i\theta}{U^\alpha}^\dagger A^\sigma {U^\alpha}
\end{eqnarray}
where $R^\alpha = e^{-i\pi\sum_iS_i^\alpha}$,  $\alpha\in\{x, y, z\}$, and $A^\sigma$ are the MPS matrices \cite{Schollwock10}. In a spin chain the $U^\alpha$ form a projective representation with $U^xU^z=e^{i\phi}U^zU^x$.  In the topological (Haldane) phase $\phi=\pi$ whereas in the trivial phase  $\phi=0$ \cite{PollmanIdent}. In the Hubbard model we find that the $U^\alpha$ do not form a closed algebra.
This is due to the fact the there is a mixture of integer and half-integer representations in the entanglement spectrum because of the charge fluctuations. 
In the Haldane phase of spin-one models the edge spins form an $SU(2)$ algebra, i.e., they are genuine spin-1/2 particles.  This shows that the edge states in the Hubbard model are importantly different from those in  pure spin models. 


In spin-one models the Haldane phase is symmetry protected by any one of three symmetries:  dihedral  ($D_2\cong Z_2\times Z_2$), time reversal and (bond) inversion symmetry  \cite{Pollmann,Chen10}. Charge fluctuations mean that time reversal and the dihedral group may not protect the Haldane phase in fermionic systems   \cite{Pollmann}. However, the Hubbard model on the triangular necklace lattice is symmetric under inversion about the bonds connecting neighboring molecules. Pollmann {\it et al.} \cite{Pollmann} have argued that this symmetry protects the Haldane phase even in fermionic systems, meaning that there must be a (quantum) phase transition  between it and a topologically trivial phase. 

Neither string order nor spin-1/2 edge states are required signatures of the Haldane phase \cite{Perez}.
Nevertheless the entanglement spectrum, {\it i.e.}, the eigenvalues of the reduced density matrix after tracing out half of the system, may only have  even-fold degeneracies in the Haldane phase  \cite{Pollmann}. Thus, the degeneracy of the entanglement  spectrum (Fig. \ref{fig:entanglement}c) confirms that the insulating phase remains topologically non-trivial even for small $U/t_c$ and large $t/t_c$. 

Finally we stress the consistency of the above picture with experiment. Llusar \emph{et al}. have shown that the magnetic susceptibility indicates the presence of doped triplets in the Mo$_3$S$_7$ units, consistent with $S \lesssim 1$ as found in our Hubbard model. No spin gap is observed down to 2 K (the lowest temperature studied)  \cite{jacs}, which is consistent with the very small spin gap  found above (cf. Fig. \ref{figure:spin_gap}b). To further test our predictions one could replace Mo$_3$S$_7$(dmit)$_3$  by ${\cal S}=1/2$ or nonmagnetic impurities \cite{Glarum};  ESR \cite{Affleck}, NMR \cite{Tedoldi} or $\mu$SR could then be 
used to search for edge excitations, which would provide a signature of SPT order. Furthermore, the expected finite energy magnon excitations of momentum $k=\pi$  in the Haldane phase  \cite{White} could be observed via neutron scattering.


We thank Matt Davis, Andrew Doherty, Carlos G\'omez-Garc\'ia, Jure Kokalj, Rosa Llusar, Ross McKenzie, Oleg Sushkov, Tom Stace, and Tony Wright for helpful conversations. This work was supported by the Australian Research Council (grants DP0878523, DP1093224,  LE120100181, and FT130100161) and MINECO (MAT2012-37263-C02-01).


%



\begin{thebibliography}{100}


 

%
%
%
%
%


%
%
%
%


\bibitem{Mott}
N. F. Mott, 
Proc. Phys. Soc. London A {\bf62}, 416 (1949). 


\bibitem{Lee}
P. A.  Lee, N. Nagaosa, and X.-G.  Wen,   
Rev. Mod. Phys. {\bf78}, 17 (2006). 

\bibitem{Anderson} 
P. W.  Anderson, 
Science {\bf235}, 1196 (1987).


\bibitem{Powell}
B. J. Powell and R. H. McKenzie,
Rep. Prog. Phys. {\bf74}, 056501 (2011).

\bibitem{Zaanen}
  J. Zaanen, G. A. Sawatzky,  and  J. W. Allen,  
  Phys. Rev. Lett. {\bf55}, 418 (1985). 
  
\bibitem{Sarma}
D. D. Sarma, J. Solid State Chem {\bf88}, 45 (1990).

\bibitem{Merino-interplay}
J. Merino, B. J. Powell and R. H. McKenzie, Phys. Rev. B {\bf 79} 161103(R) (2009).



 \bibitem{jacs} 
 R. Llusar, {\it et al}., 
 J. Am. Chem. Soc. \textbf{126}, 12076 (2004).
 
\bibitem{dft}
A. C. Jacko {\it et al}., unpublished.



\bibitem{Haldane} 
F. D. M.  Haldane, 
Phys. Rev. Lett. {\bf50}, 1153 (1983).

\bibitem{AKLT}
 I. Affleck,  T. Kennedy, E. H.  Lieb, and H.  Tasaki, 
 Phys. Rev. Lett. {\bf59}, 799 (1987).

\bibitem{Wen12}
X. Chen,  Z.-C. Gu,  Z.-X. Liu,  and X.-G.  Wen, 
Science {\bf338}, 1604 (2012).


\bibitem{GuWen}
Z.-C. Gu  and X.-G. Wen, 
Phys. Rev. B {\bf80}, 155131 (2009).


\bibitem{Pollmann}
F. Pollmann, E. Berg,  A. M.  Turner, and M.  Oshikawa, 
Phys. Rev. B {\bf85}, 075125 (2012).

\bibitem{Anfuso}
F. Anfuso  and  A. Rosch, 
Phys. Rev. B {\bf75}, 144420 (2007).

\bibitem{Sakai}
T. Sakai, M Sato, K. Okunishi, Y. Otsuka, K. Okatmoto, and C. Itoi, Phys. Rev. B {\bf78} 184415 (2008).

\bibitem{Schollwock10} U.  Schollw\"{o}ck, 
Ann. Phys. \textbf{326}, 96 (2011).

\bibitem{Kennedy}
T.  Kennedy and H.  Tasaki, 
Phys. Rev. B {\bf45}, 304 (1992). 











 





\bibitem{pert}
C. Janani, J. Merino, I. P. McCulloch and B. J. Powell, Phys. Rev. B {\bf90}, 035120 (2014).

\bibitem{Nagaoka} 
Y.  Nagaoka, 
Phys. Rev.  {\bf147}, 392 (1966). 

\bibitem{Merino}
J. Merino,  B. J.  Powell, and R. H. McKenzie,  
Phys. Rev. B {\bf73}, 235107  (2006).



\bibitem{Daul}
S. Daul and R. M. Noack, Phys. Rev. B {\bf58}, 2635 (1998).

\bibitem{Arita}
R. Arita, K. Kusakabe, K. Kuroki, and H Aoki, {\bf58}, R11833 (1998).

\bibitem{Nakano}
H. Nakano and Y. Takahashi, J. Phys. Soc. Japan {\bf72} 1191 (2003).


\bibitem{foot-coef}
Here $\varepsilon_0=0$, $a_0=-3$, $\varepsilon_1=U$, $a_1=3$, and for $n>1$
$\varepsilon_n=\frac23\left[ U+\xi\cos\left(\{\phi+2\pi n\}/3 \right)\right]$  and $a_n=2|\alpha_n|^2+|\beta_n|^2+1$, 
where $\xi =  \sqrt{U^2 + 27 t_c^2 }$,   
$\phi= \pi+\arccos[(U/\xi)^3]$,
 $\alpha_n =[-12 t_c^2 U-9 t_c^2 \varepsilon_n+U^2 \varepsilon_n-2 U \varepsilon_n^2+\varepsilon_n^3]/[\sqrt{2} (U-\varepsilon_n) (3 t_c+U-\varepsilon_n) \varepsilon_n]$,
and
$\beta_n=[U-3 t_c -\varepsilon_n ]/[U+3 t_c  -\varepsilon_n]$.  For a  derivation of this result see \cite{pert}.


\bibitem{Chen10}
X. Chen,  Z.-C. Gu, and X.-G. Wen, 
Phys. Rev. B {\bf82}, 155138 (2010).







\bibitem{Affleck} 
M. Hagiwara, K.  Katsumata,   I. Affleck,   B. I. Halperin, and J. P. Renard, 
Phys. Rev. Lett. {\bf 65}, 3181 (1990). 


\bibitem{Glarum} 
S. H. Glarum, S. Geschwind, K. M. Lee, M. L. Kaplan, and J. Michel, 
Phys. Rev. Lett. {\bf 67}, 1615 (1991). 


\bibitem{Tedoldi} 
F. Tedoldi, S. Santachiara, and M. Horvatic, 
Phys. Rev. Lett. {\bf 83},  412 (1999). 


\bibitem{White} 
S. R. White and D. A.  Huse, 
Phys. Rev. B {\bf48}, 3844 (1993). 

%

\bibitem{PollmanIdent}
F. Pollmann and  A. M.  Turner, Phys. Rev. B {\bf86}, 125441 (2012).


\bibitem{Perez}
D. P\'erez-Garc\'ia,   M. M. Wolf,    M. Sanz,   F. Verstraete, and   J. I. Cirac, 
Phys. Rev. Lett. {\bf100}, 167202 (2008).




 




 
\end{thebibliography}
\end{document}